\documentclass[conference]{IEEEtran}
\IEEEoverridecommandlockouts
\usepackage{cite}
\usepackage{amsmath,amssymb,amsfonts}
\usepackage{algorithmic}
\usepackage{graphicx}
\usepackage{textcomp}
\usepackage{xcolor}
\usepackage{setspace}
\usepackage{graphicx}
\usepackage{amsthm}
\usepackage{lscape}
\usepackage{latexsym}
\usepackage{bm}
\usepackage{epstopdf}
\usepackage{bbm}
\usepackage{cite}
\usepackage{url}
\usepackage{color}
\usepackage[caption=false,font=footnotesize]{subfig}
\usepackage{graphicx}
\usepackage{lmodern}
\usepackage[euler]{textgreek}

\usepackage{pgfplots}
\usepackage{pgfplotstable}
\newcounter{marknumber}
\pgfplotsset{
    error bars/every nth mark/.style={
        /pgfplots/error bars/draw error bar/.prefix code={
            \pgfmathtruncatemacro\marknumbercheck{mod(floor(\themarknumber/2),#1)}
            \ifnum\marknumbercheck=0
            \else
                \begin{scope}[opacity=0]
            \fi
        },
/pgfplots/error bars/draw error bar/.append code={
            \ifnum\marknumbercheck=0
            \else
                \end{scope}
            \fi
            \stepcounter{marknumber}    
        }
    }
}
\def\BibTeX{{\rm B\kern-.05em{\sc i\kern-.025em b}\kern-.08em
    T\kern-.1667em\lower.7ex\hbox{E}\kern-.125emX}}
\begin{document}

\title{DeepJSCC-Q: Channel Input Constrained \\Deep Joint Source-Channel Coding}
\author{Tze-Yang Tung, David Burth Kurka, Mikolaj Jankowski, Deniz Gündüz\\
Department of Electrical and Electronics Engineering, Imperial College London\\
\{tze-yang.tung14, d.kurka, mikolaj.jankowski17, d.gunduz\}@imperial.ac.uk
\thanks{This work was supported by the European Research Council (ERC) through project BEACON (No. 677854).}}


\maketitle
\vspace{-0.8cm}

\begin{abstract}
Recent works have shown that the task of wireless transmission of images can be learned with the use of machine learning techniques. 
Very promising results in end-to-end image quality, superior to popular digital schemes that utilize source and channel coding separation, have been demonstrated through the training of an autoencoder, with a non-trainable channel layer in the middle. 
However, these methods assume that any complex value can be transmitted over the channel, which can prevent the application of the algorithm in scenarios where the hardware or protocol can only admit certain sets of channel inputs, such as the use of a digital constellation.
Herein, we propose \emph{DeepJSCC-Q}, an end-to-end optimized joint source-channel coding scheme for wireless image transmission, which is able to operate with a fixed channel input alphabet. 
We show that \emph{DeepJSCC-Q} can achieve similar performance to models that use continuous-valued channel input.
Importantly, it preserves the graceful degradation of image quality observed in prior work when channel conditions worsen, making \emph{DeepJSCC-Q} much more attractive for deployment in practical systems.
\end{abstract}


\section{Introduction}
\label{sec:intro}

Source coding and channel coding are two essential steps in modern data transmission.
The former reduces the redundancy within source data, preserving only the essential information needed to achieve a prescribed reconstruction fidelity. 
For example, in image transmission, commonly used source coding schemes are JPEG2000 and BPG. 
Channel coding, on the other hand, introduces structured redundancy within the data to allow reliable decoding under the presence of channel noises.
It was proven by Shannon that, the separation of source and channel coding is without loss of optimality when the blocklength goes to infinity \cite{Shannon:1948}. 
Nevertheless, in practical applications we are limited to finite blocklengths, and it has been known that such communication systems can benefit from combining the two coding steps, motivating the design of joint source-channel coding (JSCC) schemes. 
Despite their potential advantages, JSCC schemes have found limited use in practice due to their lack of modularity and difficulty in designing such codes.
Recently, a method called DeepJSCC, which employs deep neural networks (DNNs) to perform a mapping from the input image space directly to channel input symbols in a joint manner was introduced in \cite{Eirina:TCCN:19}. 
It shows appealing properties, such as better or similar performance compared with state-of-the-art digital compression schemes \cite{Kurka:IZS2020}, flexibility to adapt to different source or channel models \cite{Eirina:TCCN:19,Kurka:IZS2020,yang_deep_2021}, ability to exploit channel feedback \cite{Kurka:deepjsccf:jsait}, and capability to produce adaptive-bandwidth transmission schemes \cite{Kurka:BandwidthAgile:TWComm2021}.
Importantly, \emph{graceful degradation} of image quality with respect to decreasing channel quality means that DeepJSCC is able to avoid the \textit{cliff-effect} that all separation-based schemes suffer from, where the image becomes un-decodable due to the channel quality falling below what the channel code anticipates.

One of the strengths of DeepJSCC is the fact that it learns a communication scheme from scratch, defining all transformations in a data-driven manner. 
This simplifies the code design procedure, and allows adaptation to any particular source or channel domain.
We note that the DeepJSCC approach not only combines source and channel coding into one single mapping, but it also removes the constellation diagrams used in digital schemes. 
Instead, in DeepJSCC, the encoder can transmit arbitrary complex-valued channel inputs (within the power constraint).
This can hinder the suitability of using DeepJSCC in current commercial hardware and standardized protocols, which are constrained to produce fixed sets of symbols.

In this work, we investigate the effects of constraining the channel input alphabet to a predefined constellation imposed externally. 
This constraint can be crucial for the adoption of DeepJSCC in commercially available hardware (e.g. radio transmitters), where modulators are hard-coded, limiting the output space available for the encoder, or even in incorporating such techniques into established standards such as cellular communications. 
Therefore, in this paper, we introduce a new strategy for JSCC of images, \emph{DeepJSCC-Q}, which allows for the transmission of the content through fixed pre-defined constellations. 
A series of experiments demonstrate that \emph{DeepJSCC-Q} is able to:
\begin{itemize}
    \item Achieve similar performance results to unconstrained DeepJSCC, even when using a highly constrained channel input representation.
    \item Achieve superior performance when compared to capacity-achieving low density parity check (LDPC) codes \cite{gallager_low-density_1962} with BPG compression \cite{Bellard:BPG} under the same channel input constraint.
    \item Create coherent mapping on neighboring constellation points, thus avoiding the \textit{cliff-effect} present in separation-based schemes.
\end{itemize}

\begin{figure*}
\begin{center}
\includegraphics[width=0.94\textwidth]{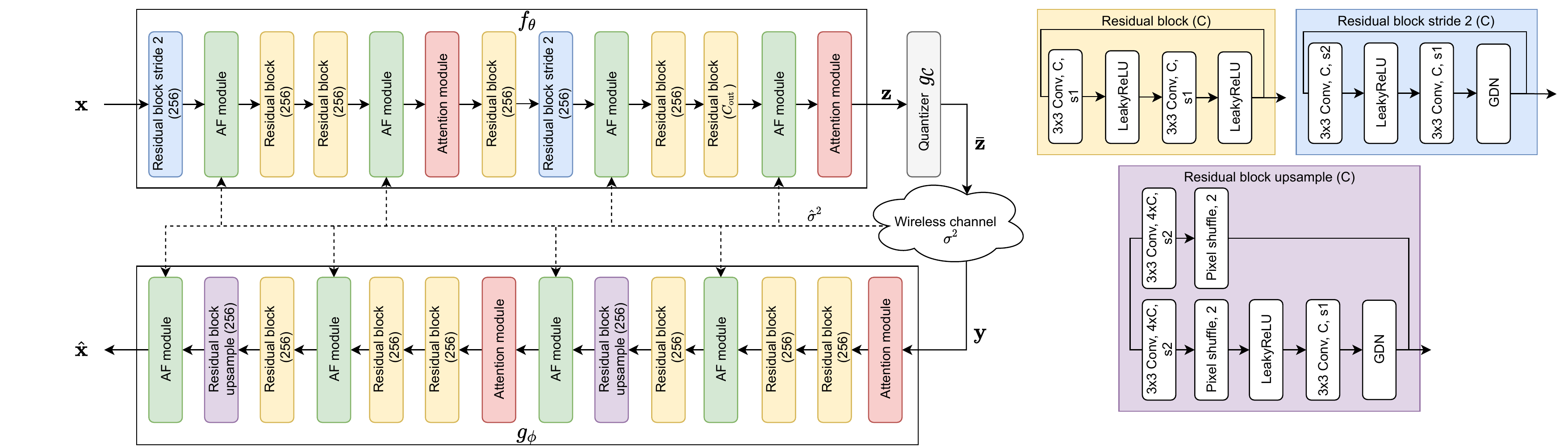}
\end{center}
  \caption{Architecture of the proposed encoder and decoder models.}
\label{fig:architecture}
\vspace{-0.4cm}
\end{figure*}

\vspace{-0.2cm}

\section{Related Works}
\label{sec:related_works}

JSCC for wireless image transmission using DNNs was first demonstrated in \cite{Eirina:TCCN:19}, and dubbed \emph{DeepJSCC}. 
The authors were able to show that by setting up the encoder and decoder in an autoencoder configuration, with a non-trainable channel layer in between, the DNN encoder was able to learn a function which maps images to continuous channel inputs, and vice versa at the decoder. 
They demonstrated that the resultant JSCC encoder and decoder was able to surpass the performance of JPEG2000 \cite{christopoulos_jpeg2000_2000} compression followed by LDPC codes \cite{gallager_low-density_1962} for channel coding.
Importantly, they showed that such schemes can avoid the \textit{cliff-effect}, exhibited by all separation-based schemes, which is when the channel quality deteriorates below the minimum channel quality to allow successful decoding, leading to a cliff-edge drop-off in the end-to-end performance.
Since then, various works have extended this result to further demonstrate the ability to exploit channel feedback \cite{Kurka:deepjsccf:jsait} and adapt to various bandwidth requirements without retraining \cite{Kurka:BandwidthAgile:TWComm2021}.
In \cite{yang_deep_2021}, the viability of DeepJSCC in an orthogonal frequency division multiplexing (OFDM) system is shown and in \cite{ding_snr-adaptive_2021} it is shown that it can be extended to multi-user scenarios with the same decoder.

However, an implicit assumption in all of the works above is the ability for the communication hardware to transmit arbitrary complex-valued channel inputs.
This may not be feasible as many commercially available hardware may have hard-coded standard protocols, making these methods less viable for real world deployment.
In \cite{choi_necst_2018}, image transmission problem is considered over a discrete input channel, where the input representation is learned using a variational autoencoder (VAE) assuming a Bernoulli prior.
They consider the transmission of MNIST images \cite{deng_mnist_2012} over a binary erasure channel (BEC) and showed that their scheme performs better than a VAE that only performs compression paired with an LDPC code.
In \cite{stark_joint_2019}, the channel coding problem with a fixed channel input alphabet is considered, and it is shown that the probability distribution of the constellation points can be learned by optimizing the bit error rate for a given channel signal-to-noise ratio (SNR).
In contrast to these works, we investigate the transmission of natural images over an additive white Gaussian noise (AWGN) channel with a finite channel input alphabet. 

\section{Problem Statement}
\label{sec:problem_def}

Herein, we consider the problem of wireless transmission of images over an additive white Gaussian noise (AWGN) channel, in which communication should be performed using a finite set of symbols.
An input image $\mathbf{x} \in \mathbb{R}^{H\times W\times C}$ (where $H$, $W$ and $C$ represent the image's height, width and color channels, respectively) is mapped with a non-linear encoder function $f_{\boldsymbol{\theta}}:\mathbb{R}^{H\times W\times C}\mapsto \mathbb{C}^k$ into a latent vector $\mathbf{z}=f_{\boldsymbol{\theta}}(\mathbf{x},\hat{\sigma}^2)$, where $\hat{\sigma}^2$ is the estimated channel noise power. 
The latent vector $\mathbf{z}$ is then mapped to a channel input alphabet $\mathcal{C}$ via the modulator $q_{\mathcal{C}}:\mathbb{C}^k\mapsto\mathbb{C}^k$, which we will define in the sequel. 
There are $M=|\mathcal{C}|$ possible symbols, and we refer to $\mathcal{C}$ as the constellation, denoted by $\mathcal{C}=\{c_1,...,c_M\}$. 
That is, $\bar{\mathbf{z}}=q_{\mathcal{C}}(\mathbf{z})$, $\bar z_i \in \mathcal{C}$, where $\bar z_i$ is the $i$th element of the channel input vector $\bar{\mathbf{z}}$.
We define the average power of the constellation assuming equally likely inputs as
\begin{equation}
    P = \frac{1}{M}\sum_{j=1}^M |c_j|^2.
\end{equation}
The channel input $\bar{\mathbf{z}}$ is transmitted through an AWGN channel, producing the channel output $\mathbf{y} = \bar{\mathbf{z}} + \boldsymbol{\eta}$, where $\boldsymbol{\eta}\sim CN(0,\sigma^2I_{k\times k})$ is a complex Gaussian vector with covariance matrix $\sigma^2 I_{k\times k}$, and $I_{k\times k}$ is the $k\times k$ identity matrix. 
Finally, a receiver passes the channel output through a non-linear decoder function $g_{\boldsymbol{\phi}}:\mathbb{C}^k\mapsto\mathbb{R}^{H\times W\times C}$ to produce a reconstruction of the input $\hat{\mathbf{x}}=g_{\boldsymbol{\phi}}(\mathbf{y},\hat{\sigma}^2)$.

We assume that both the transmitter and the receiver can estimate the channel noise power, denoted by $\hat{\sigma}^2$, in order to select the best resource allocation strategy.
We define the channel SNR as
\begin{equation}
    \text{SNR} = 10\log_{10}\Big(\frac{P}{{\sigma}^2}\Big) \text{ dB},
\end{equation}
and the SNR estimated by the encoder and decoder as
\begin{equation}
    \text{SNR}_{\text{Est}} = 10\log_{10}\Big(\frac{P}{\hat{\sigma}^2}\Big) \text{ dB}.
\end{equation}
We also define the \emph{bandwidth compression ratio} as
\begin{equation}
    \rho = \frac{k}{H\times W\times C}~\text{channel symbols/pixel},
\end{equation}
where a smaller number reflects more compression.

To this end, we propose \emph{DeepJSCC-Q}, a DNN model and an end-to-end training strategy, which learns the encoder and decoder parameters $\boldsymbol{\theta}, \boldsymbol{\phi}$, and effectively utilize the available symbols in order to recover the transmitted images with satisfactory quality. 
We will consider the use of quantization to represent our modulator $q_\mathcal{C}$, such that each quantization level represents a point in the constellation. 
As in previous works \cite{Kurka:IZS2020,Eirina:TCCN:19}, we utilize an autoencoder architecture to jointly train the encoder and the decoder.
However, one of the drawbacks of the previous works was the need to train multiple networks, one for each channel condition.
To address this issue, \cite{xu_wireless_2020} proposed an attention feature (AF) module, motivated by resource assignment strategies in traditional JSCC schemes \cite{sayood_joint_2000}, which allows the network to learn to assign different weights to different features for a given SNR.
This is done by deliberately randomizing the channel SNR during training and providing the AF modules the current SNR, such that $\hat{\sigma}^2=\sigma^2$.
By doing so, the results in \cite{xu_wireless_2020} show that the single model performs at least as well as the models trained for each SNR individually.
We adopt the AF module proposed by \cite{xu_wireless_2020} in \emph{DeepJSCC-Q} to obtain a single model that can work over a range of SNRs given an estimate of the channel condition.

We propose a fully convolutional encoder and decoder architecture as shown in Fig. \ref{fig:architecture}. 
In the architecture, $C$ refers to the number of channels in the output tensor of the convolution operation.
$C_{\text{out}}$ refers to the number of channels in the final output tensor of the encoder $f_{\boldsymbol{\theta}}$, which controls the number of channel uses $k$ per image.
The "Pixel shuffle" module, within the "Residual block upsample" module, is used to increases the height and width of the input tensor dimensions by reshaping the input tensor, such that the channel dimension is reduced while the height and width dimensions are increased. 
This was first proposed in \cite{shi_real-time_2016} as a less computationally expensive method for increasing the CNN tensor dimensions without requiring large number of parameters, like transpose convolutional layers.
The GDN layer refers to generalized divisive normalization, initially proposed in \cite{balle2015density}, and shown to be effective in density modeling and compression of images.
The Attention layer refers to the simplified attention module proposed in \cite{cheng_learned_2020}, which reduces the computation cost of the attention module originally proposed in \cite{wang_non-local_2018}.
While the attention layer have been used in \cite{cheng_learned_2020} and \cite{wang_non-local_2018} to improve the compression efficiency by learning to focus on image regions that require higher bit rate, in our model it is used to allow adaptive allocation of channel bandwidth and power resources.

\subsection{Quantization}
\label{subs:quantization}

\begin{figure}
\centering
\includegraphics[width=0.95\linewidth]{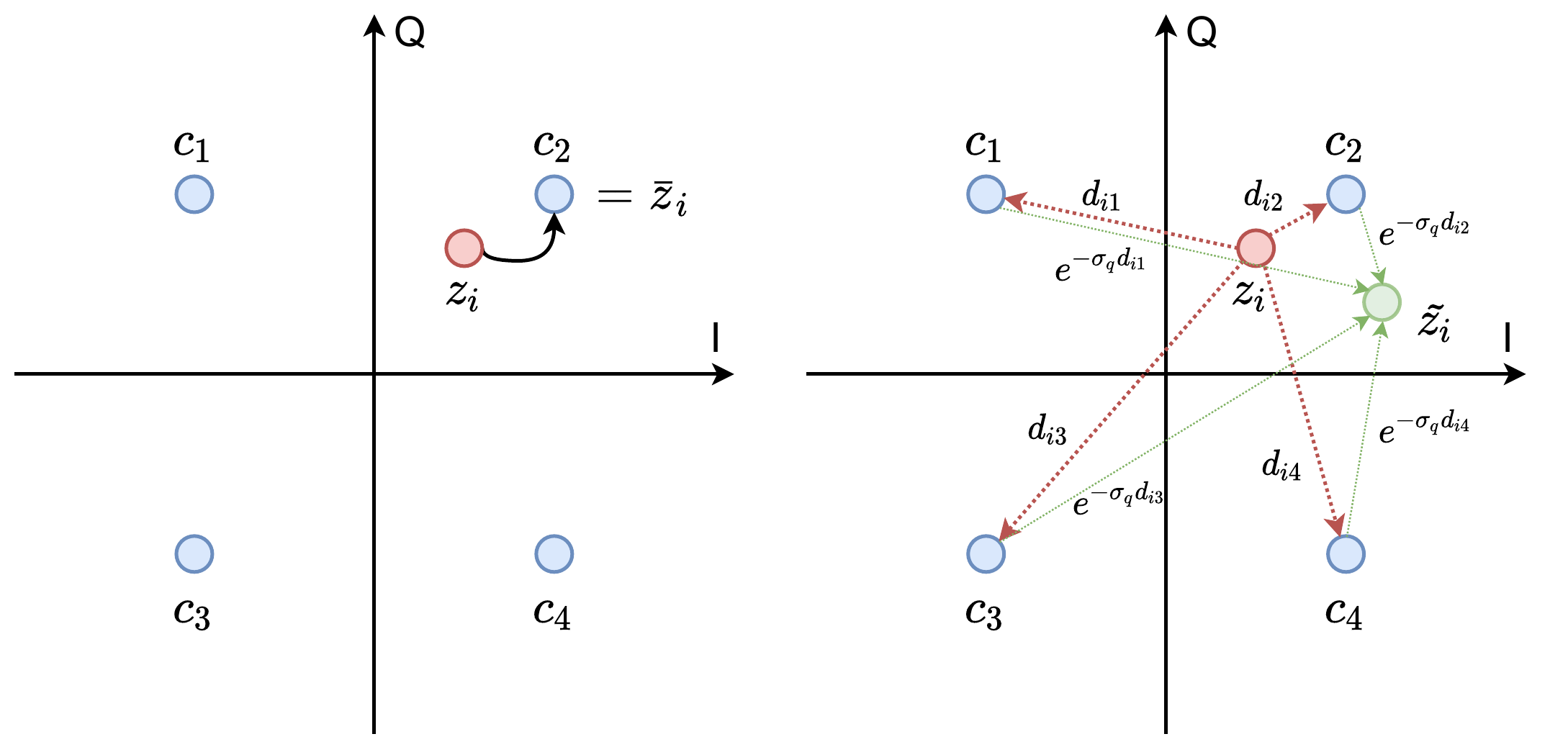}
  \caption{Illustration of the soft-to-hard quantization procedure for a single value $z_i$ using a QPSK constellation.
  The hard quatized value $\bar{z}_i$ (left) simply maps the latent value $z_i$ to the nearest point in the constellation, while the soft quantized value $\tilde{z}_i$ (right) is the softmax weighted sum of the constellation points according to $l_2$ distance.}
\label{fig:soft_hard_quant}
\vspace{-0.8cm}
\end{figure}

In order to produce an encoder that outputs a fixed constellation, we perform quantization of the latent vector generated by the encoder, $\bar{\mathbf{z}}=g_\mathcal{C}(\mathbf{z})$. 
Given the encoder output $\mathbf{z}$, we first obtain a ``hard" quantization, which simply maps element $z_i\in\mathbf{z}$ to the nearest symbol in $\mathcal{C}$.
This forms the channel input $\bar{\mathbf{z}}$.
However, this operation is not differentiable.
In order to obtain a differentiable approximation of the hard quantization operation, we will use the  ``soft" quantization approach, proposed in \cite{Agustsson:softQuant:NIPS2017}. 
In this approach, each quantized symbol is generated as the softmax weighted sum of the symbols in $\mathcal{C}$ based on their distances from $z_i$; 
that is,
\begin{equation}
\label{eq:soft_quantization}
    \tilde{z_i} = \sum_{j=1}^M \frac{e^{-\sigma_q d_{ij}}}{\sum_{n=1}^M e^{-\sigma_q d_{in}}} c_j,
\end{equation}
where $\sigma_q$ is a parameter controlling the ``hardness'' of the assignment, and $d_{ij} = ||z_i-c_j||^2_2$, is $l_2$ distance between the latent value $z_i$ and the constellation point $c_j$.
As such, in the forward pass, the quantizer uses the hard quantization, corresponding to the channel input $\bar{\mathbf{z}}$, and in the backward pass, the gradient from the soft quantization $\tilde{\mathbf{z}}$ is used to update $\boldsymbol{\theta}$.
That is,
\begin{equation}
    \frac{\partial\bar{\mathbf{z}}}{\partial \mathbf{z}} =
    \frac{\partial\tilde{\mathbf{z}}}{\partial \mathbf{z}}.
\end{equation}
A diagram illustrating the soft-to-hard quantizer for 4-QAM, also known as the quadrature phase shift keying (QPSK), is shown in Fig. \ref{fig:soft_hard_quant}.

We consider constellation symbols that are uniformly distributed in a square lattice over the complex plane, similar to QAM-modulation. 
For QAM constellation consisting of $M$ symbols, denoted as M-QAM, we define the max amplitude 
$
    A_{max} = \frac{(M-1)}{2}\sqrt{\frac{12P}{(M^2-1)}},
$ 
and inter symbol distance 
$
    d_{sym} = \sqrt{\frac{12P}{(M^2-1)}},
$ 
where $P$ is the average power of the constellation under uniform distribution, i.e., $\mathbb{E}[\mathcal{C}^2] = \frac{1}{M} \sum_{i=1}^M |{c}_{i}|^2 = P$.

\subsection{Training Strategy}
\label{subsec:training}

Given that our model is end-to-end optimized, we are free to choose any distortion metric.
Two common distortion metrics for images are the mean squared error (MSE),
\begin{equation}
    \text{MSE}(\mathbf{x},\hat{\mathbf{x}})=||\mathbf{x}-\hat{\mathbf{x}}||_2^2,
    \label{eq:mse_loss}
\end{equation}
and the structural similarity index (SSIM),
\begin{equation}
    \text{SSIM}(\mathbf{x},\hat{\mathbf{x}})=
    \left(\frac{2\mu_{\mathbf{x}}\mu_{\hat{\mathbf{x}}}+v_1}{\mu_{\mathbf{x}}^2+\mu_{\hat{\mathbf{x}}}^2+v_1}\right) 
    \left(\frac{2\sigma_{\mathbf{x}}\sigma_{\hat{\mathbf{x}}}+v_2}{\sigma_{\mathbf{x}}^2+\sigma_{\hat{\mathbf{x}}}^2+v_2}\right),
    \label{eq:ssim_loss}
\end{equation}
where $\mu_{\mathbf{x}}$, $\sigma^2_{\mathbf{x}}$, $\sigma^2_{\mathbf{x}\hat{\mathbf{x}}}$ are the mean and variance of $\mathbf{x}$, and the covariance between $\mathbf{x}$ and $\hat{\mathbf{x}}$, respectively, and $v_1$, $v_2$ are coefficients for numeric stability.
When used on an RGB image, SSIM is computed for all three color channels and then averaged.
Note that a higher SSIM value indicates a lower distortion, with a maximum of 1, therefore when used as a training loss, $1-\text{SSIM}$ is used.

A common complimentary metric to the MSE distortion to measure image reconstruction quality is the peak signal-to-noise ratio (PSNR) defined as
\begin{equation}
    \text{PSNR}(\mathbf{x},\hat{\mathbf{x}})=\log_{10}\bigg(\frac{A^2}{\text{MSE}(\mathbf{x},\hat{\mathbf{x}})}\bigg)~\text{dB},
    \label{eq:psnr_def}
\end{equation}
where $A$ is the maximum possible value for a given pixel. 
For a 24 bit RGB pixel, $A=255$.

\begin{figure} 
    \centering
  \subfloat[PSNR\label{subfig:psnr_soft_hard_graceful}]{%
    \begin{tikzpicture}
        \pgfplotsset{
            legend style={
                font=\fontsize{4}{4}\selectfont,
                at={(1.0,.0)},
                anchor=south east,
            },
            width=0.5\textwidth,
            height=0.37\textwidth,
            xmin=0,
            xmax=10,
            ymin=20,
            ymax=35,
            xtick distance=2,
            ytick distance=2,
            xlabel={SNR (dB)},
            ylabel={PSNR (dB)},
            grid=both,
            grid style={line width=.1pt, draw=gray!10},
            major grid style={line width=.2pt,draw=gray!50},
            every axis/.append style={
                x label style={
                    font=\fontsize{8}{8}\selectfont,
                    at={(axis description cs:0.5,-0.0)},
                    },
                y label style={
                    font=\fontsize{8}{8}\selectfont,
                    at={(axis description cs:0.03,0.5)},
                    },
                x tick label style={
                    font=\fontsize{8}{8}\selectfont,
                    /pgf/number format/.cd,
                    fixed,
                    fixed zerofill,
                    precision=0,
                    /tikz/.cd
                    },
                y tick label style={
                    font=\fontsize{8}{8}\selectfont,
                    /pgf/number format/.cd,
                    fixed,
                    fixed zerofill,
                    precision=1,
                    /tikz/.cd
                    },
            }
        }
        \begin{axis}
        \addplot[green, solid, line width=0.9pt, mark=*, mark options={fill=green, scale=1.1}, error bars/.cd, y dir=both, y explicit, every nth mark=1] 
                table [x=snr, y=4qam_soft_hard_KL(snr2), 
                y error=4qam_soft_hard_KL(std), col sep=comma]
                {data/4qam_tsnr2_psnr.csv};
        \addlegendentry{\textit{DeepJSCC-Q} QPSK 
        ($\lambda=0.05$, $\text{SNR}_{\text{Est}}=2dB$)}
        
        \addplot[red, solid, line width=0.9pt, mark=triangle*, mark options={fill=red, scale=1.1}, error bars/.cd, y dir=both, y explicit, every nth mark=1] 
                table [x=snr, y=4qam_soft_hard_KL(snr6), 
                y error=4qam_soft_hard_KL(std), col sep=comma]
                {data/4qam_tsnr6_psnr.csv};
        \addlegendentry{\textit{DeepJSCC-Q} QPSK 
        ($\lambda=0.05$, $\text{SNR}_{\text{Est}}=6dB$)}
        
        \addplot[blue, solid, line width=0.9pt, mark=square*, mark options={fill=blue, scale=1.1, solid}, error bars/.cd, y dir=both, y explicit, every nth mark=1] 
                table [x=snr, y=16qam_soft_hard(snr8), 
                y error=16qam_soft_hard(std), col sep=comma]
                {data/16qam_tsnr8_psnr.csv};
        \addlegendentry{\textit{DeepJSCC-Q} 16-QAM 
        ($\lambda=0.05$, $\text{SNR}_{\text{Est}}=8dB$)}
        
        
        \addplot[color=black, dashed, line width=1.2pt, mark=*, mark options={fill=black, solid, scale=1.1}, 
        error bars/.cd, y dir=both, y explicit, every nth mark=1] 
        table [x=SNR, y=0.5_ldpc_psnr, 
        y error=0.5_ldpc_psnr(std), col sep=comma]
        {data/bpg_4qam_psnr.csv};
        \addlegendentry{BPG + LDPC 1/2 QPSK}
        
        \addplot[color=black, dashed, line width=1.2pt, mark=triangle*, mark options={fill=black, solid, scale=1.1}, 
        error bars/.cd, y dir=both, y explicit, every nth mark=1] 
        table [x=SNR, y=0.66_ldpc_psnr, 
        y error=0.66_ldpc_psnr(std), col sep=comma]
        {data/bpg_4qam_psnr.csv};
        \addlegendentry{BPG + LDPC 2/3 QPSK}
        
        \addplot[color=black, dashed, line width=1.2pt, mark=square*, mark options={fill=black, solid, scale=1.1}, 
        error bars/.cd, y dir=both, y explicit, every nth mark=1] 
        table [x=SNR, y=0.5_ldpc_psnr, 
        y error=0.5_ldpc_psnr(std), col sep=comma]
        {data/bpg_16qam_psnr.csv};
        \addlegendentry{BPG + LDPC 1/2 16-QAM}
        \end{axis}
        \end{tikzpicture}
    }
    \\
    \vspace{-0.4cm}
  \subfloat[SSIM\label{subfig:ssim_soft_hard_graceful}]{%
    \begin{tikzpicture}
        \pgfplotsset{
            legend style={
                font=\fontsize{4}{4}\selectfont,
                at={(1.0,0.)},
                anchor=south east,
            },
            width=0.5\textwidth,
            height=0.37\textwidth,
            xmin=0,
            xmax=10,
            ymin=0.77,
            ymax=1,
            xtick distance=2,
            ytick distance=0.05,
            xlabel={SNR (dB)},
            ylabel={SSIM},
            grid=both,
            grid style={line width=.1pt, draw=gray!10},
            major grid style={line width=.2pt,draw=gray!50},
            every axis/.append style={
                x label style={
                    font=\fontsize{8}{8}\selectfont,
                    at={(axis description cs:0.5, -0.0)},
                    },
                y label style={
                    font=\fontsize{8}{8}\selectfont,
                    at={(axis description cs:0.03,0.5)},
                    },
                x tick label style={
                    font=\fontsize{8}{8}\selectfont,
                    /pgf/number format/.cd,
                    fixed,
                    fixed zerofill,
                    precision=0,
                    /tikz/.cd
                    },
                y tick label style={
                    font=\fontsize{8}{8}\selectfont,
                    /pgf/number format/.cd,
                    fixed,
                    fixed zerofill,
                    precision=2,
                    /tikz/.cd
                    },
            }
        }
        \begin{axis}
        \addplot[green, solid, line width=0.9pt, mark=*, mark options={fill=green, scale=1.1}, error bars/.cd, y dir=both, y explicit, every nth mark=1] 
                table [x=snr, y=4qam_soft_hard_KL(snr2), y
                error=4qam_soft_hard_KL(std), col sep=comma]
                {data/4qam_tsnr2_ssim.csv};
        \addlegendentry{\textit{DeepJSCC-Q} QPSK 
        ($\lambda=0.05$, $\text{SNR}_{\text{Est}}=2dB$)}
        
        \addplot[red, solid, line width=0.9pt, mark=triangle*, mark options={fill=red, scale=1.1}, error bars/.cd, y dir=both, y explicit, every nth mark=1] 
                table [x=snr, y=4qam_soft_hard_KL(snr6), 
                y error=4qam_soft_hard_KL(std), col sep=comma]
                {data/4qam_tsnr6_ssim.csv};
        \addlegendentry{\textit{DeepJSCC-Q} QPSK 
        ($\lambda=0.05$, $\text{SNR}_{\text{Est}}=6dB$)}
        
        \addplot[blue, solid, line width=0.9pt, mark=square*, mark 
        options={fill=blue, scale=1.1, solid}, error bars/.cd, y dir=both, y explicit, every nth mark=1] 
                table [x=snr, y=16qam_soft_hard(snr8), 
                y error=16qam_soft_hard(std), col sep=comma]
                {data/16qam_tsnr8_ssim.csv};
        \addlegendentry{\textit{DeepJSCC-Q} 16-QAM 
        ($\lambda=0.05$, $\text{SNR}_{\text{Est}}=8dB$)}
        
        
        \addplot[color=black, dashed, line width=1.2pt, mark=*, mark options={fill=black, solid, scale=1.1}, 
        error bars/.cd, y dir=both, y explicit, every nth mark=1] 
        table [x=SNR, y=0.5_ldpc_ssim, 
        y error=0.5_ldpc_ssim(std), col sep=comma]
        {data/bpg_4qam_ssim.csv};
        \addlegendentry{BPG + LDPC 1/2 QPSK}
        
        \addplot[color=black, dashed, line width=1.2pt, mark=triangle*, mark options={fill=black, solid, scale=1.1}, 
        error bars/.cd, y dir=both, y explicit, every nth mark=1] 
        table [x=SNR, y=0.66_ldpc_ssim, 
        y error=0.66_ldpc_ssim(std), col sep=comma]
        {data/bpg_4qam_ssim.csv};
        \addlegendentry{BPG + LDPC 2/3 QPSK}
        
        \addplot[color=black, dashed, line width=1.2pt, mark=square*, mark options={fill=black, solid, scale=1.1}, 
        error bars/.cd, y dir=both, y explicit, every nth mark=1] 
        table [x=SNR, y=0.5_ldpc_ssim, 
        y error=0.5_ldpc_ssim(std), col sep=comma]
        {data/bpg_16qam_ssim.csv};
        \addlegendentry{BPG + LDPC 1/2 16-QAM}
        \end{axis}
        \end{tikzpicture}
        }
  \caption{
  Effect of channel estimation error. Here, we compare \emph{DeepJSCC-Q} using different $\text{SNR}_{\text{Est}}$ and distortion metrics to BPG using LDPC codes.
  }
  \label{fig:soft_hard_graceful} 
  \vspace{-0.6cm}
\end{figure}
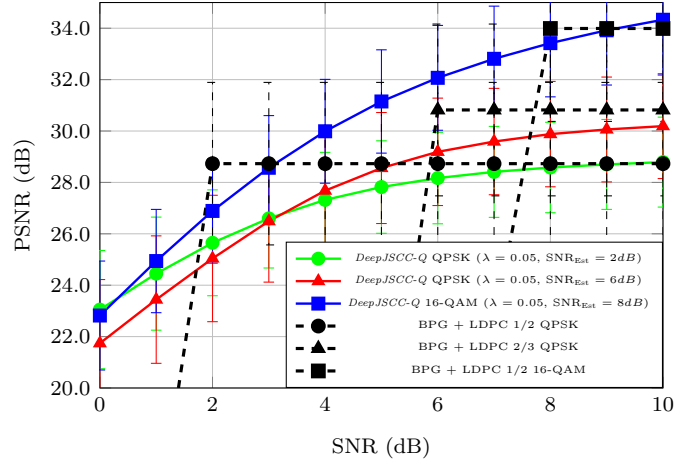
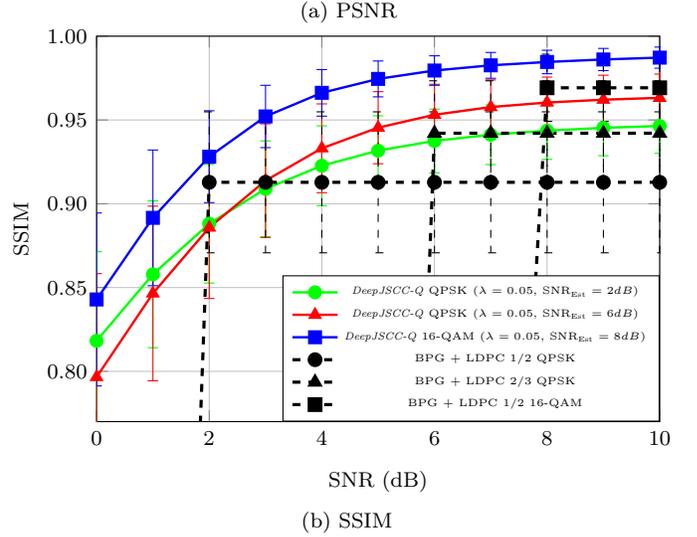

In order to utilize the available constellation points in the most optimal way and encourage the transmitted signal to have an average power close to 1, we introduce a regularization term based on the Kullback-Leibler (KL) divergence between the distribution $P(\mathcal{C}|\mathbf{z})$ and a uniform distribution over the constellation set $\mathcal{U}(\mathcal{C})$. 
The KL divergence between distributions $P_W$ and $P_V$ is defined as
\begin{equation}
    D_{\text{KL}}(P_W\ ||\ P_V) =
    \mathbb{E}\left[\log\left(\frac{P_W}{P_V}\right)\right],
\end{equation}
and it measures how different the two distributions are, with $D_{\text{KL}}(P_W\ ||\ P_V)=0 \Longleftrightarrow P_W=P_V$. 
The distribution $P(\mathcal{C}|\mathbf{z})$ represents the probability of selecting a point in the constellation set $\mathcal{C}$ given the encoded latent vector $\mathbf{z}$.
By regularizing the distortion loss with the KL divergence $D_{\text{KL}}(P(\mathcal{C}|\mathbf{z})\ ||\ \mathcal{U}(\mathcal{C}))$, we encourage the quantizer $q_\mathcal{C}$ to explore the available constellation points and thus force the transmitted signal power to be as close to 1 as possible.
To model the distribution $P(\mathcal{C}|\mathbf{z})$, we utilize the softmax weights used in the soft assignment in Eq. (\ref{eq:soft_quantization}).
Since the weights sum to 1, we can treat them as probabilities and average the probability of each constellation point over the training batch to obtain an estimate of $P(\mathcal{C}|\mathbf{z})$. 
That is, the probability of selecting a constellation point $c_j$ given a batch of latent vectors $\{\mathbf{z}^v\}_{v=1}^{|\mathcal{B}|}$, where $\mathcal{B}$ is a batch of encoded latent vectors, can be estimated as 

\begin{equation}
    \hat{P}(c_j|\mathbf{z}) =
    \frac{1}{|\mathcal{B}|k}\sum_{v=1}^{|\mathcal{B}|}\sum_{i=1}^{k}
    \frac{e^{-\sigma_q d_{ij}^{v}}}{\sum_{n=1}^M e^{-\sigma_q d_{in}^{v}}},
\end{equation}
where $d_{ij}^v=||z_i^v - c_j||_2^2$ is the $l_2$ distance between the constellation point $c_j$ and the $i$th element in the $v$th latent vector in the batch.
Therefore, the final loss function we use for training is:
\begin{equation}
\label{eq:loss}
    l(\mathbf{x},\hat{\mathbf{x}}) = d(\mathbf{x}, \hat{\mathbf{x}}) + \lambda D_{\text{KL}}(\hat{P}(\mathcal{C}|\mathbf{z})\ ||\ \mathcal{U}(\mathcal{C})),
\end{equation}
where $d(\cdot,\cdot)$ is the distortion metric (either MSE or $1-\text{SSIM}$) and $\lambda$ is the weighting parameter to control the amount of regularization.

\section{Experimental Results}
\label{sec:results}

\begin{figure} 
    \centering
  \subfloat[PSNR\label{subfig:psnr_soft_hard_mod}]{%
    \begin{tikzpicture}
        \pgfplotsset{
            legend style={
                font=\fontsize{4}{4}\selectfont,
                at={(1.0,.0)},
                anchor=south east,
            },
            width=0.5\textwidth,
            xmin=-2,
            xmax=10,
            ymin=18,
            ymax=38,
            xtick distance=2,
            ytick distance=2,
            xlabel={SNR (dB)},
            ylabel={PSNR (dB)},
            grid=both,
            grid style={line width=.1pt, draw=gray!10},
            major grid style={line width=.2pt,draw=gray!50},
            every axis/.append style={
                x label style={
                    font=\fontsize{8}{8}\selectfont,
                    at={(axis description cs:0.5,-0.0)},
                    },
                y label style={
                    font=\fontsize{8}{8}\selectfont,
                    at={(axis description cs:0.03,0.5)},
                    },
                x tick label style={
                    font=\fontsize{8}{8}\selectfont,
                    /pgf/number format/.cd,
                    fixed,
                    fixed zerofill,
                    precision=0,
                    /tikz/.cd
                    },
                y tick label style={
                    font=\fontsize{8}{8}\selectfont,
                    /pgf/number format/.cd,
                    fixed,
                    fixed zerofill,
                    precision=1,
                    /tikz/.cd
                    },
            }
        }
        \begin{axis}
        \addplot[green, solid, line width=0.9pt, mark=*, mark options={fill=green, scale=1.1}, error bars/.cd, y dir=both, y explicit, every nth mark=1] 
                table [x=snr, y=4qam_soft_hard_KL, y
                error=4qam_soft_hard_KL(std), col sep=comma]
                {data/jscc_4qam_psnr.csv};
        \addlegendentry{\textit{DeepJSCC-Q} QPSK 
        ($\lambda=0.05$, $\text{SNR}_{\text{Est}}=\text{SNR}$)}
        
        \addplot[red, solid, line width=0.9pt, mark=triangle*, mark options={fill=red, scale=1.1}, error bars/.cd, y dir=both, y explicit, every nth mark=1] 
                table [x=snr, y=16qam_soft_hard_KL, 
                y error=16qam_soft_hard_KL(std), col sep=comma]
                {data/jscc_16qam_psnr.csv};
        \addlegendentry{\textit{DeepJSCC-Q} 16-QAM 
        ($\lambda=0.05$, $\text{SNR}_{\text{Est}}=\text{SNR}$)}
        
        \addplot[blue, solid, line width=0.9pt, mark=square*, mark options={fill=blue, scale=1.1, solid}, error bars/.cd, y dir=both, y explicit, every nth mark=1] 
                table [x=snr, y=64qam_soft_hard, 
                y error=64qam_soft_hard(std), col sep=comma]
                {data/jscc_64qam_psnr.csv};
        \addlegendentry{\textit{DeepJSCC-Q} 64-QAM 
        ($\lambda=0.00$, $\text{SNR}_{\text{Est}}=\text{SNR}$)}
        
        \addplot[color=black, dashed, line width=1.2pt, mark=*, mark options={fill=black, solid, scale=1.1}, 
        error bars/.cd, y dir=both, y explicit, every nth mark=1] 
        table [x=SNR, y=0.33_ldpc_psnr, 
        y error=0.33_ldpc_psnr(std), col sep=comma]
        {data/bpg_4qam_psnr.csv};
        \addlegendentry{BPG + LDPC 1/3 QPSK}
        
        \addplot[color=cyan, dashed, line width=1.2pt, mark=triangle*, mark options={fill=cyan, solid, scale=1.1}, 
        error bars/.cd, y dir=both, y explicit, every nth mark=1] 
        table [x=SNR, y=0.5_ldpc_psnr, 
        y error=0.5_ldpc_psnr(std), col sep=comma]
        {data/bpg_4qam_psnr.csv};
        \addlegendentry{BPG + LDPC 1/2 QPSK}
        
        \addplot[color=black, dashed, line width=1.2pt, mark=triangle*, mark options={fill=black, solid, scale=1.1}, 
        error bars/.cd, y dir=both, y explicit, every nth mark=1] 
        table [x=SNR, y=0.66_ldpc_psnr, 
        y error=0.66_ldpc_psnr(std), col sep=comma]
        {data/bpg_4qam_psnr.csv};
        \addlegendentry{BPG + LDPC 2/3 QPSK}
        
        \addplot[color=cyan, dashed, line width=1.2pt, mark=square*, mark options={fill=cyan, solid, scale=1.1}, 
        error bars/.cd, y dir=both, y explicit, every nth mark=1] 
        table [x=SNR, y=0.5_ldpc_psnr, 
        y error=0.5_ldpc_psnr(std), col sep=comma]
        {data/bpg_16qam_psnr.csv};
        \addlegendentry{BPG + LDPC 1/2 16-QAM}
        
        \end{axis}
        \end{tikzpicture}
    }
    \\
    \vspace{-0.4cm}
  \subfloat[SSIM\label{subfig:ssim_soft_hard_mod}]{%
    \begin{tikzpicture}
        \pgfplotsset{
            legend style={
                font=\fontsize{4}{4}\selectfont,
                at={(1.0,0.)},
                anchor=south east,
            },
            height=0.39\textwidth,
            width=0.5\textwidth,
            xmin=-2,
            xmax=10,
            ymin=0.7,
            ymax=1,
            xtick distance=2,
            ytick distance=0.05,
            xlabel={SNR (dB)},
            ylabel={SSIM},
            grid=both,
            grid style={line width=.1pt, draw=gray!10},
            major grid style={line width=.2pt,draw=gray!50},
            every axis/.append style={
                x label style={
                    font=\fontsize{8}{8}\selectfont,
                    at={(axis description cs:0.5, -0.0)},
                    },
                y label style={
                    font=\fontsize{8}{8}\selectfont,
                    at={(axis description cs:0.03,0.5)},
                    },
                x tick label style={
                    font=\fontsize{8}{8}\selectfont,
                    /pgf/number format/.cd,
                    fixed,
                    fixed zerofill,
                    precision=0,
                    /tikz/.cd
                    },
                y tick label style={
                    font=\fontsize{8}{8}\selectfont,
                    /pgf/number format/.cd,
                    fixed,
                    fixed zerofill,
                    precision=2,
                    /tikz/.cd
                    },
            }
        }
        \begin{axis}
        \addplot[green, solid, line width=0.9pt, mark=*, mark options={fill=green, scale=1.1}, error bars/.cd, y dir=both, y explicit, every nth mark=1] 
                table [x=snr, y=4qam_soft_hard_KL, y
                error=4qam_soft_hard_KL(std), col sep=comma]
                {data/jscc_4qam_ssim.csv};
        \addlegendentry{\textit{DeepJSCC-Q} QPSK 
        ($\lambda=0.05$, $\text{SNR}_{\text{Est}}=\text{SNR}$)}
        
        \addplot[red, solid, line width=0.9pt, mark=triangle*, mark options={fill=red, scale=1.1}, error bars/.cd, y dir=both, y explicit, every nth mark=1] 
                table [x=snr, y=16qam_soft_hard_KL, 
                y error=16qam_soft_hard_KL(std), col sep=comma]
                {data/jscc_16qam_ssim.csv};
        \addlegendentry{\textit{DeepJSCC-Q} 16-QAM 
        ($\lambda=0.05$, $\text{SNR}_{\text{Est}}=\text{SNR}$)}
        
        \addplot[blue, solid, line width=0.9pt, mark=square*, mark options={fill=blue, scale=1.1, solid}, error bars/.cd, y dir=both, y explicit, every nth mark=1] 
                table [x=snr, y=64qam_soft_hard, 
                y error=64qam_soft_hard(std), col sep=comma]
                {data/jscc_64qam_ssim.csv};
        \addlegendentry{\textit{DeepJSCC-Q} 64-QAM 
        ($\lambda=0.00$, $\text{SNR}_{\text{Est}}=\text{SNR}$)}
        
        \addplot[color=black, dashed, line width=1.2pt, mark=*, mark options={fill=black, solid, scale=1.1}, 
        error bars/.cd, y dir=both, y explicit, every nth mark=1] 
        table [x=SNR, y=0.33_ldpc_ssim, 
        y error=0.33_ldpc_ssim(std), col sep=comma]
        {data/bpg_4qam_ssim.csv};
        \addlegendentry{BPG + LDPC 1/3 QPSK}
        
        \addplot[color=cyan, dashed, line width=1.2pt, mark=triangle*, mark options={fill=cyan, solid, scale=1.1}, 
        error bars/.cd, y dir=both, y explicit, every nth mark=1] 
        table [x=SNR, y=0.5_ldpc_ssim, 
        y error=0.5_ldpc_ssim(std), col sep=comma]
        {data/bpg_16qam_ssim.csv};
        \addlegendentry{BPG + LDPC 1/2 QPSK}
        
        \addplot[color=black, dashed, line width=1.2pt, mark=triangle*, mark options={fill=black, solid, scale=1.1}, 
        error bars/.cd, y dir=both, y explicit, every nth mark=1] 
        table [x=SNR, y=0.66_ldpc_ssim, 
        y error=0.66_ldpc_ssim(std), col sep=comma]
        {data/bpg_4qam_ssim.csv};
        \addlegendentry{BPG + LDPC 2/3 QPSK}
        
        \addplot[color=cyan, dashed, line width=1.2pt, mark=square*, mark options={fill=cyan, solid, scale=1.1}, 
        error bars/.cd, y dir=both, y explicit, every nth mark=1] 
        table [x=SNR, y=0.5_ldpc_ssim, 
        y error=0.5_ldpc_ssim(std), col sep=comma]
        {data/bpg_16qam_ssim.csv};
        \addlegendentry{BPG + LDPC 1/2 16-QAM}
        
        \end{axis}
        \end{tikzpicture}
        }
  \caption{
  Comparison of \emph{DeepJSCC-Q} for $\text{SNR}_{\text{Est}}=\text{SNR}$ to BPG using LDPC codes.
  }
  \label{fig:soft_hard_mod} 
\vspace{-0.8cm}
\end{figure}
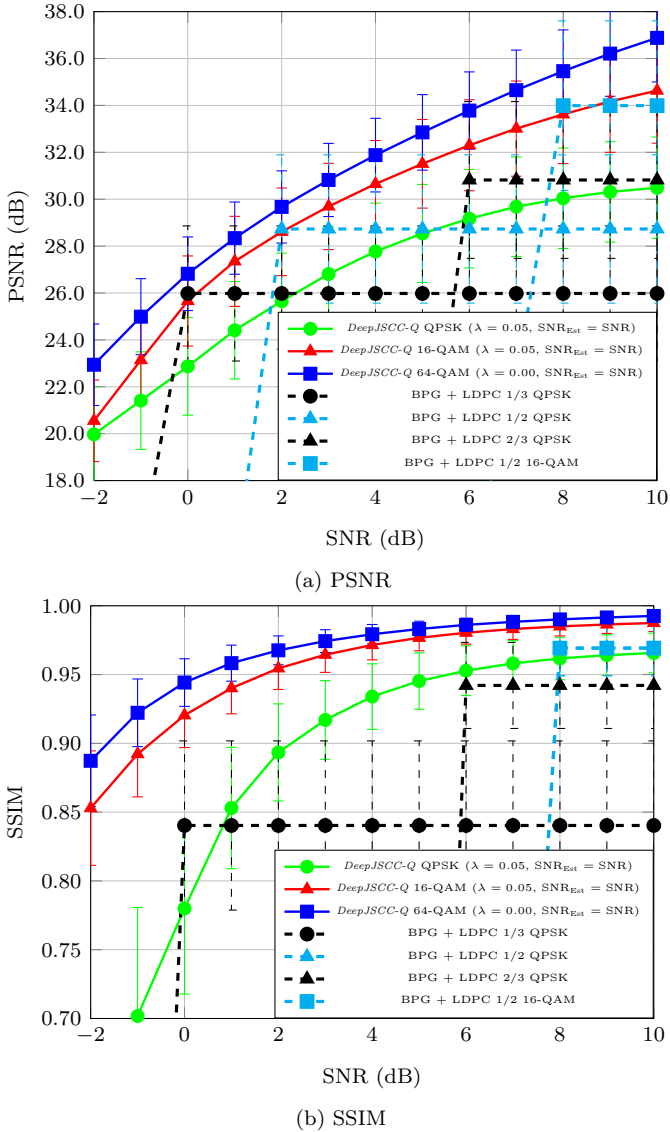

In this section we perform a series of experiments to demonstrate the performance of \emph{DeepJSCC-Q}. 
Herein, we consider the CIFAR10 dataset \cite{Krizhevsky:CIFAR} which consists of 60000 RGB images of $32\times 32$ resolution. 
We split the dataset into $5:1$ for training and testing, respectively.
We use the Pytorch \cite{paszke_automatic_2017} library and Adam \cite{kingma_adam_2017} optimizer with learning rate $0.0001$, $\beta_1=0.9$, and $\beta_2=0.999$ to train our encoder and decoder networks.
We use a batch size of 32 and early stopping with a patience of 8 epochs, where the maximum number of training epochs is 1000.
We implement learning rate scheduling, where the learning rate is reduced by a factor of 0.8 if the loss does not improve for 4 epochs in a row.
We use an average constellation power $P=1$ and change the channel noise power $\sigma^2$ accordingly to obtain any given SNR. 
During training, the channel SNR is drawn uniformly randomly $\text{SNR}\sim\mathcal{U}(0,10)$.
The softmax hardness assignment parameter is chosen as $\sigma_q=100$.
We set the weighting for the regularizer $\lambda=0.05$ when the size of the constellation $M$ is less than 64, as we experimentally found it to be helpful to encourage the channel input to be more uniformly distributed across the constellation set, while for larger constellations, $\lambda=0$ performed better, indicating that it is more beneficial to choose a subset of available symbols with higher probability than using all symbols with the same frequency.
We use $C_{\text{out}}=40$ for all models, which corresponds to $\rho=0.4166$.


To compare the performance of our solution, we consider separation-based schemes, in which images are first compressed using the BPG \cite{Bellard:BPG} codec, before an LDPC code \cite{gallager_low-density_1962} is used as the channel code.
We compare the average image quality over the test dataset, with error bars showing the standard deviation of the image quality metric.
Fig. \ref{fig:soft_hard_graceful} shows the effects of channel estimation error on the performance of \emph{DeepJSCC-Q}.
The constellation and $\text{SNR}_{\text{Est}}$ values used for \emph{DeepJSCC-Q} are chosen to coincide with the SNR values at which the separation schemes would fail, in order to highlight the behavior of \emph{DeepJSCC-Q} around the cliff edges of the separation-based schemes. 
Note that this is done by fixing the channel noise estimate $\hat{\sigma}^2$, while varying the actual channel noise power $\sigma^2$.
For all M-QAM modulation orders shown here, \emph{DeepJSCC-Q} exhibited graceful degradation of image quality with decreasing channel quality.
This is similar to the \emph{DeepJSCC} result from \cite{Eirina:TCCN:19}, but we are able to obtain the same behavior despite being constrained to a finite digital constellation.
Moreover, when compared to the separation-based results, the \emph{DeepJSCC-Q} 16-QAM model performed close to the envelope of all the separation-based schemes, with the model trained using the SSIM distortion performing considerably above the envelope.
This shows that the end-to-end optimized \emph{DeepJSCC-Q} is fundamentally different from separation-based schemes, as it is able to maintain a performance on par with separation-based schemes without the need to adapt its code rate.

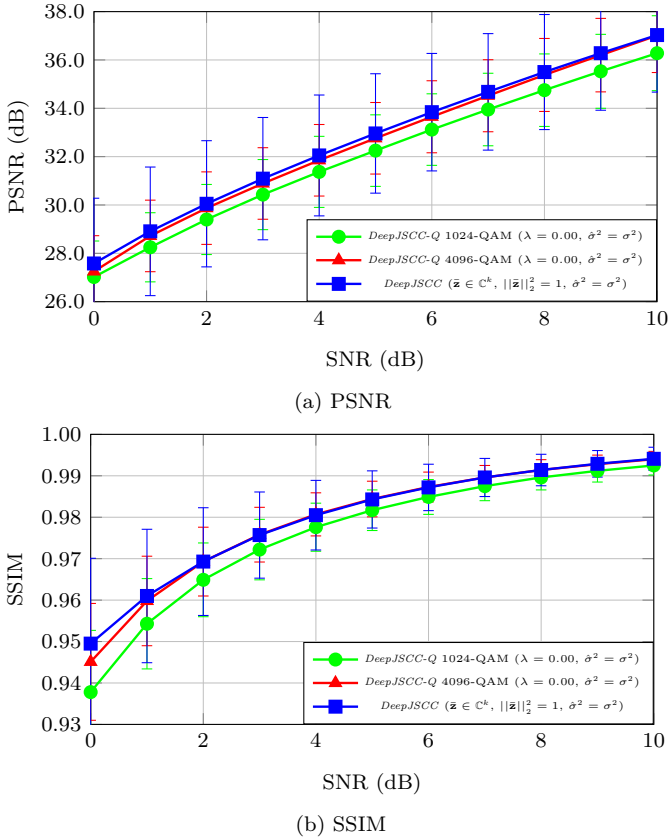
\begin{figure} 
    \centering
  \subfloat[PSNR\label{subfig:psnr_high_mod_v_jscc}]{%
    \begin{tikzpicture}
        \pgfplotsset{
            legend style={
                font=\fontsize{4}{4}\selectfont,
                at={(1.0,.0)},
                anchor=south east,
            },
            width=0.5\textwidth,
            height=0.3\textwidth,
            xmin=0,
            xmax=10,
            ymin=26,
            ymax=38,
            xtick distance=2,
            ytick distance=2,
            xlabel={SNR (dB)},
            ylabel={PSNR (dB)},
            grid=both,
            grid style={line width=.1pt, draw=gray!10},
            major grid style={line width=.2pt,draw=gray!50},
            every axis/.append style={
                x label style={
                    font=\fontsize{8}{8}\selectfont,
                    at={(axis description cs:0.5,-0.0)},
                    },
                y label style={
                    font=\fontsize{8}{8}\selectfont,
                    at={(axis description cs:0.03,0.5)},
                    },
                x tick label style={
                    font=\fontsize{8}{8}\selectfont,
                    /pgf/number format/.cd,
                    fixed,
                    fixed zerofill,
                    precision=0,
                    /tikz/.cd
                    },
                y tick label style={
                    font=\fontsize{8}{8}\selectfont,
                    /pgf/number format/.cd,
                    fixed,
                    fixed zerofill,
                    precision=1,
                    /tikz/.cd
                    },
            }
        }
        \begin{axis}
        \addplot[green, solid, line width=0.9pt, mark=*, mark options={fill=green, scale=1.1}, error bars/.cd, y dir=both, y explicit, every nth mark=1] 
                table [x=snr, y=1024qam_soft_hard, y
                error=1024qam_soft_hard(std), col sep=comma]
                {data/jscc_1024qam_psnr.csv};
        \addlegendentry{\textit{DeepJSCC-Q} 1024-QAM 
        ($\lambda=0.00$, $\hat{\sigma}^2=\sigma^2$)}
        
        \addplot[red, solid, line width=0.9pt, mark=triangle*, mark options={fill=red, scale=1.1}, error bars/.cd, y dir=both, y explicit, every nth mark=1] 
                table [x=snr, y=4096qam_soft_hard, 
                y error=4096qam_soft_hard(std), col sep=comma]
                {data/jscc_4096qam_psnr.csv};
        \addlegendentry{\textit{DeepJSCC-Q} 4096-QAM 
        ($\lambda=0.00$, $\hat{\sigma}^2=\sigma^2$)}
        
        \addplot[blue, solid, line width=0.9pt, mark=square*, mark options={fill=blue, scale=1.1, solid}, error bars/.cd, y dir=both, y explicit, every nth mark=1] 
                table [x=snr, y=PSNR, 
                y error=PSNR(std), col sep=comma]
                {data/jscc_continuous.csv};
        \addlegendentry{\textit{DeepJSCC}  
        ($\bar{\mathbf{z}}\in\mathbb{C}^k$, $||\bar{\mathbf{z}}||_2^2=1$, $\hat{\sigma}^2=\sigma^2$)}
        \end{axis}
        \end{tikzpicture}
    }
    \\
    \vspace{-0.3cm}
  \subfloat[SSIM\label{subfig:ssim_high_mod_v_jscc}]{%
    \begin{tikzpicture}
        \pgfplotsset{
            legend style={
                font=\fontsize{4}{4}\selectfont,
                at={(1.0,0.)},
                anchor=south east,
            },
            width=0.5\textwidth,
            height=0.3\textwidth,
            xmin=0,
            xmax=10,
            ymin=0.93,
            ymax=1,
            xtick distance=2,
            ytick distance=0.01,
            xlabel={SNR (dB)},
            ylabel={SSIM},
            grid=both,
            grid style={line width=.1pt, draw=gray!10},
            major grid style={line width=.2pt,draw=gray!50},
            every axis/.append style={
                x label style={
                    font=\fontsize{8}{8}\selectfont,
                    at={(axis description cs:0.5, -0.0)},
                    },
                y label style={
                    font=\fontsize{8}{8}\selectfont,
                    at={(axis description cs:0.03,0.5)},
                    },
                x tick label style={
                    font=\fontsize{8}{8}\selectfont,
                    /pgf/number format/.cd,
                    fixed,
                    fixed zerofill,
                    precision=0,
                    /tikz/.cd
                    },
                y tick label style={
                    font=\fontsize{8}{8}\selectfont,
                    /pgf/number format/.cd,
                    fixed,
                    fixed zerofill,
                    precision=2,
                    /tikz/.cd
                    },
            }
        }
        \begin{axis}
        \addplot[green, solid, line width=0.9pt, mark=*, mark options={fill=green, scale=1.1}, error bars/.cd, y dir=both, y explicit, every nth mark=1] 
                table [x=snr, y=1024qam_soft_hard, y
                error=1024qam_soft_hard(std), col sep=comma]
                {data/jscc_1024qam_ssim.csv};
        \addlegendentry{\textit{DeepJSCC-Q} 1024-QAM 
        ($\lambda=0.00$, $\hat{\sigma}^2=\sigma^2$)}
        
        \addplot[red, solid, line width=0.9pt, mark=triangle*, mark options={fill=red, scale=1.1}, error bars/.cd, y dir=both, y explicit, every nth mark=1] 
                table [x=snr, y=4096qam_soft_hard_KL, 
                y error=4096qam_soft_hard_KL(std), col sep=comma]
                {data/jscc_4096qam_ssim.csv};
        \addlegendentry{\textit{DeepJSCC-Q} 4096-QAM 
        ($\lambda=0.00$, $\hat{\sigma}^2=\sigma^2$)}
        
        \addplot[blue, solid, line width=0.9pt, mark=square*, mark options={fill=blue, scale=1.1, solid}, error bars/.cd, y dir=both, y explicit, every nth mark=1] 
                table [x=snr, y=SSIM, 
                y error=SSIM(std), col sep=comma]
                {data/jscc_continuous.csv};
        \addlegendentry{\textit{DeepJSCC}  
        ($\bar{\mathbf{z}}\in\mathbb{C}^k$, $||\bar{\mathbf{z}}||_2^2=1$, 
        $\hat{\sigma}^2=\sigma^2$)}
        \end{axis}
        \end{tikzpicture}
        }
  \caption{
  Comparison of \emph{DeepJSCC-Q} using high modulation orders to \emph{DeepJSCC} with a continuous channel input, as in \cite{Eirina:TCCN:19}.
  }
  \label{fig:high_mod_v_jscc} 
  \vspace{-0.4cm}
\end{figure}

The reason \emph{DeepJSCC-Q} performs this way is related to the quantization error.
A higher order modulation essentially corresponds to greater number of quantization levels of the encoder output $\mathbf{z}$, and thus lower quantization error of $\bar{\mathbf{z}}$ in representing $\mathbf{z}$.
Since DeepJSCC has already been shown to surpass the performance of BPG and LDPC codes by \cite{Kurka:IZS2020}, naturally as we increase the constellation order $M$, the performance of \emph{DeepJSCC-Q} will approach that of \emph{DeepJSCC}.
This is also why \emph{DeepJSCC-Q} with 16-QAM constellation performs better than both QPSK models in Fig. \ref{fig:soft_hard_graceful}, except at the low SNR regime, although that is due to the $\text{SNR}_{\text{Est}}$ used for the 16-QAM model being much greater than the channel SNR in that region.
In Fig. \ref{fig:soft_hard_mod}, where the $\text{SNR}_{\text{Est}}=\text{SNR}$, we can see that not only does the 16-QAM \emph{DeepJSCC-Q} model perform better than the QPSK model, the 64-QAM model also performs better than either of those models, for all channel SNRs tested.
As further evidence that increasing the modulation order of \emph{DeepJSCC-Q} leads to the performance of \emph{DeepJSCC-Q} approaching that of \emph{DeepJSCC} \cite{Eirina:TCCN:19}, we investigate very high order modulations $M=1024,4096$, as shown in Fig. \ref{fig:high_mod_v_jscc}.
To compare to DeepJSCC, we simply remove the quantizer $q_\mathcal{C}$ and normalize the output power of the encoder, such that the channel input is
\begin{equation}
    \bar{\mathbf{z}} = \frac{\sqrt{kP}}{||\mathbf{z}||_2}\mathbf{z}.
\end{equation}
From Fig. \ref{fig:high_mod_v_jscc}, the performance of \emph{DeepJSCC-Q} approaches \emph{DeepJSCC} as the modulation order increases, with $M=4096$ the two models performing almost the same.
\vspace{-0.2cm}

\section{Conclusions}
\label{sec:conclusions}

In this paper, we have proposed \emph{DeepJSCC-Q}, an end-to-end optimized JSCC scheme for image transmission that is able to utilize a fixed modulation constellation and achieve similar performance to unquantized \emph{DeepJSCC}, as previously proposed by \cite{Eirina:TCCN:19}.
Even with such a constraint, we are able to achieve superior performance to separation-based schemes using BPG for source coding and LDPC for channel coding, all the while avoiding the \textit{cliff-effect} that plagues the separation-based schemes.
We also show that with sufficiently high modulation order, \emph{DeepJSCC-Q} can approach the performance of \emph{DeepJSCC}, which does not have a fixed channel input constellation.
As such, if such constellations are available on the hardware, \emph{DeepJSCC-Q} can perform nearly as well as \emph{DeepJSCC}.
This makes the viability of \emph{DeepJSCC-Q} in existing commercial hardware with standardized protocols much more attractive.

\bibliographystyle{ieeetr}
\bibliography{references.bib}

\begin{thebibliography}{10}

\bibitem{Shannon:1948}
C.~E. Shannon, ``A mathematical theory of communication,'' {\em Bell Syst.
  Tech. J.}, vol.~27, pp.~379--423 and 623--656, July and October 1948.

\bibitem{Eirina:TCCN:19}
E.~{Bourtsoulatze}, D.~{Burth Kurka}, and D.~{Gündüz}, ``Deep joint
  source-channel coding for wireless image transmission,'' {\em IEEE
  Transactions on Cognitive Communications and Networking}, vol.~5,
  pp.~567--579, Sep. 2019.

\bibitem{Kurka:IZS2020}
D.~Burth~Kurka and D.~G{\"u}nd{\"u}z, ``Joint source-channel coding of images
  with (not very) deep learning,'' in {\em International Zurich Seminar on
  Information and Communication (IZS 2020). Proceedings}, pp.~90--94, ETH
  Zurich, 2020.

\bibitem{yang_deep_2021}
M.~Yang, C.~Bian, and H.-S. Kim, ``Deep {joint} {source} {channel} {coding} for
  {wireless image} {transmission} with {OFDM},'' {\em arXiv:2101.03909 [cs,
  eess, math]}, May 2021.
\newblock arXiv: 2101.03909.

\bibitem{Kurka:deepjsccf:jsait}
D.~B. Kurka and D.~Gündüz, ``Deepjscc-f: Deep joint source-channel coding of
  images with feedback,'' {\em IEEE Journal on Selected Areas in Information
  Theory}, vol.~1, no.~1, pp.~178--193, 2020.

\bibitem{Kurka:BandwidthAgile:TWComm2021}
D.~B. Kurka and D.~Gündüz, ``Bandwidth-agile image transmission with deep
  joint source-channel coding,'' {\em IEEE Transactions on Wireless
  Communications}, pp.~1--1, 2021.

\bibitem{gallager_low-density_1962}
R.~Gallager, ``Low-density parity-check codes,'' {\em IRE Transactions on
  Information Theory}, vol.~8, pp.~21--28, Jan. 1962.
\newblock Conference Name: IRE Transactions on Information Theory.

\bibitem{Bellard:BPG}
F.~Bellard, {\em Better Portable Graphics}, 2014 (accessed March 13, 2020).
\newblock \url{https://bellard.org/bpg/}.

\bibitem{christopoulos_jpeg2000_2000}
C.~Christopoulos, A.~Skodras, and T.~Ebrahimi, ``The {JPEG2000} still image
  coding system: an overview,'' {\em IEEE Transactions on Consumer
  Electronics}, vol.~46, pp.~1103--1127, Nov. 2000.

\bibitem{ding_snr-adaptive_2021}
M.~Ding, J.~Li, M.~Ma, and X.~Fan, ``{SNR}-{adaptive} {deep} {joint}
  {source}-{channel} {coding} for {wireless} {image} {transmission},'' in {\em
  {ICASSP} 2021 - 2021 {IEEE} {International} {Conference} on {Acoustics},
  {Speech} and {Signal} {Processing} ({ICASSP})}, pp.~1555--1559, June 2021.
\newblock ISSN: 2379-190X.

\bibitem{choi_necst_2018}
K.~Choi, K.~Tatwawadi, T.~Weissman, and S.~Ermon, ``{NECST}: {Neural} {Joint}
  {Source}-{Channel} {Coding},'' {\em International Conference on Machine
  Learning (ICML)}, 2019.

\bibitem{deng_mnist_2012}
L.~Deng, ``The {MNIST} {Database} of {Handwritten} {Digit} {Images} for
  {Machine} {Learning} {Research} [{Best} of the {Web}],'' {\em IEEE Signal
  Processing Magazine}, vol.~29, pp.~141--142, Nov. 2012.
\newblock Conference Name: IEEE Signal Processing Magazine.

\bibitem{stark_joint_2019}
M.~Stark, F.~A. Aoudia, and J.~Hoydis, ``Joint {learning} of {geometric} and
  {probabilistic} {constellation} {shaping},'' {\em IEEE Globecom Workshops},
  Dec. 2019.

\bibitem{xu_wireless_2020}
J.~Xu, B.~Ai, W.~Chen, A.~Yang, P.~Sun, and M.~Rodrigues, ``Wireless {image}
  {transmission} {using} {deep} {source} {channel} {coding} {with} {attention}
  {modules},'' {\em IEEE Transactions on Circuits and Systems for Video
  Technology}, pp.~1--1, 2021.

\bibitem{sayood_joint_2000}
K.~Sayood, H.~H. Otu, and N.~Demir, ``Joint source/channel coding for variable
  length codes,'' {\em IEEE Transactions on Communications}, vol.~48,
  pp.~787--794, May 2000.
\newblock Conference Name: IEEE Transactions on Communications.

\bibitem{shi_real-time_2016}
W.~Shi, J.~Caballero, F.~Huszár, J.~Totz, A.~P. Aitken, R.~Bishop,
  D.~Rueckert, and Z.~Wang, ``Real-{time} {single} {image} and {video}
  {super}-{resolution} {using} an {efficient} {sub}-{pixel} {convolutional}
  {neural} {network},'' pp.~1874--1883, June 2016.
\newblock ISSN: 1063-6919.

\bibitem{balle2015density}
J.~Ball{\'e}, V.~Laparra, and E.~P. Simoncelli, ``Density modeling of images
  using a generalized normalization transformation,'' {\em arXiv preprint
  arXiv:1511.06281}, 2015.

\bibitem{cheng_learned_2020}
Z.~Cheng, H.~Sun, M.~Takeuchi, and J.~Katto, ``Learned {image} {compression}
  with {discretized} {Gaussian} {mixture} {likelihoods} and {attention}
  {modules},'' {\em Conference on Computer Vision and Pattern Recognition
  (CVPR)}, pp.~7939--7948, 2020.

\bibitem{wang_non-local_2018}
X.~Wang, R.~Girshick, A.~Gupta, and K.~He, ``Non-local {neural} {networks},''
  in {\em {Conference} on {Computer} {Vision} and {Pattern} {Recognition}
  (CVPR)}, pp.~7794--7803, June 2018.
\newblock ISSN: 2575-7075.

\bibitem{Agustsson:softQuant:NIPS2017}
E.~Agustsson {\em et~al.}, ``Soft-to-hard vector quantization for end-to-end
  learning compressible representations,'' in {\em Advances in Neural
  Information Processing Systems 30}, pp.~1141--1151, Curran Associates, Inc.,
  2017.

\bibitem{Krizhevsky:CIFAR}
A.~Krizhevsky, ``Learning multiple layers of features from tiny images,'' 2009.

\bibitem{paszke_automatic_2017}
A.~Paszke {\em et~al.}, ``Pytorch: An imperative style, high-performance deep
  learning library,'' in {\em Advances in Neural Information Processing Systems
  32}, pp.~8024--8035, Curran Associates, Inc., 2019.

\bibitem{kingma_adam_2017}
D.~P. Kingma and J.~Ba, ``Adam: {A} {Method} for {Stochastic} {Optimization},''
  {\em arXiv:1412.6980 [cs]}, Jan. 2017.
\newblock arXiv: 1412.6980.

\end{thebibliography}

\end{document}